\documentclass[twocolumn,amsmath,amssymb,floatfix,nofootinbib]{revtex4}                                                

\usepackage{url}
\usepackage{epsf, array, color}
\usepackage{graphicx}
\usepackage{braket}
\usepackage[linktocpage,colorlinks,urlcolor=blue]{hyperref}

\begin{document}

\newcommand\scalemath[2]{\scalebox{#1}{\mbox{\ensuremath{\displaystyle #2}}}}

\title{Code conversion with the quantum Golay code for a universal transversal gate set}
\author{Matthew Sullivan}
\email{msullivan1@bnl.gov}
\affiliation{Department of Physics, Brookhaven National Laboratory, Upton, New York 11973 U.S.A.}
\date{\today}

\begin{abstract}
The $[[7,1,3]]$ Steane code and $[[23,1,7]]$ quantum Golay code have been identified as good candidates for fault-tolerant quantum computing via code concatenation. These two codes have transversal implementations of all Clifford gates, but require some other scheme for fault-tolerant $T$ gates. Using magic states, Clifford operations, and measurements is one common scheme, but magic state distillation can have a large overhead. Code conversion is one avenue for implementing a universal gate set fault-tolerantly without the use of magic state distillation. Analogously to how the $[[7,1,3]]$ Steane code can be fault-tolerantly converted to and from the $[[15,1,3]]$ Reed-Muller code which has a transversal $T$ gate, the $[[23,1,7]]$ Golay code can be converted to a $[[95,1,7]]$ triorthogonal code with a transversal $T$ gate. A crucial ingredient to this procedure is the $[[49,1,5]]$ triorthogonal code, which can itself be seen as related to the self-dual $[[17,1,5]]$ 2D color code. Additionally, a method for code conversion based on a transversal CNOT between the codes, rather than stabilizer measurements, is described.
\end{abstract}

\maketitle

\section{Introduction}
\label{sec:intro}
Quantum error-correcting codes enable errors on a number of individual physical qubits to be corrected while preserving the logical state. Code concatenation enables one to suppress logical error rates to any desired level, as long as the physical error rates are below some threshold\cite{shor1996fault}. Some alternative schemes to concatenation based on certain $2D$ lattices are surface codes\cite{KITAEV20032} and color codes\cite{PhysRevLett.97.180501}. Both color codes and surface codes have geometrically local stabilizers in 2 dimensions, but $2D$ color codes notably can allow for transversal implementations of all Clifford gates.  Another class of codes which has received considerable development in recent years is quantum low-density parity-check (qLDPC) codes\cite{PRXQuantum.2.040101,10.1145/3519935.3520017,10.1145/3564246.3585101,10.1109/TIT.2023.3267945}. However, general techniques for implementing the required set of logical gates are still being investigated for qLDPC codes\cite{JochymOConnor2019faulttolerantgates,Quintavalle2023partitioningqubits}. In contrast, concatenated codes are well-understood. Another useful method is bosonic codes\cite{Ofek2016ExtendingTL,Hu:2019zbe,PhysRevA.64.012310,PRXQuantum.2.020101}, which provide a method for obtaining good qubits from infinite-dimensional continuous-variable quantum systems which can be combined with any desired qubit stabilizer codes, including concatenated code schemes, to obtain an error threshold\cite{PhysRevLett.112.120504,PhysRevA.104.062434,PRXQuantum.2.020101}. Among quantum error correcting codes, the quantum Golay code has been recognized as a good candidate for code concatenation with a high threshold\cite{cross2009faulttolerance}. The quantum Golay code is a 23-qubit code, based off the classical Golay code, which encodes one logical qubit and can correct errors on up to three physical qubits.

The ability to perform arbitrary logical operations on encoded qubits is crucial for achieving fault-tolerant universal quantum computation. The Eastin-Knill theorem demonstrates that there are no quantum error-correcting codes which can implement a universal gate set transversally \cite{eastin2009restrictions}. There are codes which can implement all of the logical Clifford gates transversally, such as the 7-qubit Steane code\cite{steane1996error} and the quantum Golay code, but Clifford gates do not give universality. Using a source of noisy `magic states', and a magic state distillation procedure to increase the fidelity of said magic states, is one of the standard ways to implement the required non-Clifford gates fault-tolerantly and achieve universal quantum computation\cite{PhysRevA.71.022316}. There do exist, however, quantum codes which can implement the logical $\pi/8$ gate, also called the $T$ gate, transversally, such as those constructed from triorthogonal binary matrices\cite{bravyi2012magic}. There are also $3D$ generalizations of color codes\cite{PhysRevLett.98.160502} which allow transversal implementation of the $T$ gate while having geometrically local stabilizers in 3 dimensions.
 The Clifford gates along with the $T$ gate would suffice for a universal gate set. This leads to one of the possible methods for implementing a universal gate set fault-tolerantly: the fault-tolerant conversion between a code with transversal $T$ and a code with transversal Clifford gates.

Anderson, Duclos-Cianci, and Poulin showed how this code conversion can be done rather simply for the 7-qubit Steane code and the 15-qubit Reed-Muller code, and more generally for the family of Reed-Muller codes\cite{anderson2014fault}.  Related to this, Paetznick and Reichardt\cite{paetznick2013universal} had earlier shown how to implement logical Hadamard fault-tolerantly using only transversal gates and error correction for the 15-qubit Reed-Muller code (framed as a particular gauge choice of 6 of the logical qubits for the $[[15,7,3]]$ quantum Hamming code). In this paper, this general scheme will be discussed and then applied to the quantum Golay code. The ingredients of this scheme are
\begin{enumerate}
\item a self-dual Calderbank-Shor-Steane (CSS) code encoding 1 logical qubit which has fully transversal logical Pauli operators,
\item a triorthogonal code encoding 1 logical qubit.
\end{enumerate}
Related methods have been applied to $2D$ color codes by such names as stacked codes\cite{PhysRevA.93.022323}, gauge color codes\cite{PhysRevA.93.052332}, and doubled color codes\cite{bravyi2015doubled}. The resulting codes are not $2D$ topological codes, although Bravyi and Cross showed how they can be cleverly implemented using local operations\cite{bravyi2015doubled}. The general procedure for this construction will be referred to as `code-doubling'. This constructon will be used to give a $[[95,1,7]]$ triorthogonal code with transversal $T$ gate which can be converted to and from the quantum Golay code.

\section{General code-doubling procedure}
\label{sec:doubling}
The general procedure to construct a triorthogonal code from an $n$ qubit self-dual CSS code is similar to the construction of the Reed-Muller code from two copies of the Steane code, except that instead of a single ancillary qubit, a sequence of $m$ qubits encoding a smaller triorthogonal code is used. As shall be demonstrated, this is necessary to preserve the code distance. This procedure is the general version of the implementations discussed in Refs~\cite{PhysRevA.93.022323,PhysRevA.93.052332,bravyi2015doubled}. To more closely match the standard notation for triorthogonal codes, the odd-weight logical and the even-weight generators of the self-dual CSS code will be written as a matrix $B_{\textrm{sd}}$ in block form as
\begin{equation}
\label{eq:selfdual}
B_{\textrm{sd}} = \begin{bmatrix}B_{\textrm{sd},(1)} \\ \hline B_{\textrm{sd},(0)}\end{bmatrix}.
\end{equation}
It will be useful to write an additional matrix $E_{\textrm{sd}}$ with even-weight rows such that the rows of $E_{\textrm{sd}}$ and $B_{\textrm{sd}}$ form a complete basis of $Z_{2}^{n}$, where $n$ is the number of physical qubits in the self-dual code. For the smaller triorthogonal code, the corresponding triorthogonal matrix is similarly written in $B_{\textrm{tri}}$ in block form as
\begin{equation}
\label{eq:smalltriorthogonal}
B_{\textrm{tri}} = \begin{bmatrix}B_{\textrm{tri},(1)} \\ \hline B_{\textrm{tri},(0)}\end{bmatrix}.
\end{equation}
The orthogonal complement of $B_{\textrm{tri}}$ is spanned by the rows of the matrix 
\begin{equation}
\label{eq:tricomplement}
B_{\textrm{tri}}^\bot =
\begin{bmatrix} B_{\textrm{tri},(0)} \\ C_{\textrm{tri}}
\end{bmatrix},
\end{equation} for some matrix $C_{\textrm{tri}}$ corresponding to the additional set of $Z$ stabilizers used for the triorthogonal code.

The derived triorthogonal matrix from code-doubling will now be presented. In block form, it is given by
\begin{equation}
\label{eq:doubledcode}
B_{\textrm{doubled}} =
\begin{bmatrix}
B_{\textrm{doubled},(1)}\\ \hline
B_{\textrm{doubled},(0)}
\end{bmatrix} =
\begin{bmatrix}
B_{\textrm{sd},(1)} & B_{\textrm{sd},(1)} & B_{\textrm{tri},(1)}  \\ \hline
B_{\textrm{sd},(0)} & B_{\textrm{sd},(0)} & 0_{\textrm{tri}}  \\
0_{\textrm{sd}} & 1_{\textrm{sd}} & B_{\textrm{tri},(1)}\\
0_{\textrm{sd}} & 0_{\textrm{sd}} & B_{\textrm{tri},(0)}
\end{bmatrix} ,
\end{equation}
with $0_{\textrm{sd}}$, $0_{\textrm{tri}}$, $1_{\textrm{sd}}$ representing a 1 by $n$ submatrix containing all 0 or all 1 entries with $n$ given by the number of qubits in either the self-dual or triorthogonal code, and with $B_{\textrm{sd},(0)}$, $B_{\textrm{sd},(1)}$, $B_{\textrm{tri},(0)}$, $B_{\textrm{tri},(1)}$ as in Eqs.~\ref{eq:selfdual}~and~\ref{eq:smalltriorthogonal}. The code derived from Eq.~\ref{eq:doubledcode} will be referred to as a `doubled code'.

Before continuing, note the use of both $B_{\textrm{sd},(1)}$ and $1_{\textrm{sd}}$ in Eq~\ref{eq:doubledcode}. In Section~\ref{sec:intro}, it was mentioned that the construction would require that the self-dual CSS code only encoded a single logical qubit, and that it had fully transversal logical Pauli operators, so $B_{\textrm{sd},(1)}$ and $1_{\textrm{sd}}$ might ostensibly seem to be the same. Of course, there are smaller representations of the logical Pauli operators obtained from multiplication with stabilizers, so $B_{\textrm{sd},(1)}$ is not unique. $B_{\textrm{sd},(1)}$ is used where any smaller representation works in the construction and $1_{\textrm{sd}}$ where only the fully transversal representation works. Since $1_{\textrm{sd}}$ is necessary for one stabilizer, this means that this construction will have a large weight stabilizer when using a large self-dual code.

The reason that $B_{\textrm{doubled}}$ is a triorthogonal matrix is fairly simple: non-trivial products of three rows of $B_{\textrm{doubled}}$ involve one of the following: 
\begin{enumerate}
\item products of three rows of $B_{\textrm{tri}}$, which is itself a triorthogonal matrix;
\item twice the product of rows of $B_{\textrm{sd}}$, which gives 0 in arithmetic modulo 2;
\item the product of $1_{\textrm{sd}}$ and two rows of $B_{\textrm{sd}}$, which reduces to the product of two rows of $B_{\textrm{sd}}$, and which is 0 by the orthogonality of self-dual CSS codes.
\end{enumerate}
Note that by the same arguments, this doubled code construction can work in general to construct a $(k+1)$-orthogonal code out of a $k$-orthogonal code and a smaller $(k+1)$-orthgonal code, thus allowing conversions between many different kinds of codes in analogy to the whole Reed-Muller code family. However, only conversion between a self-dual code and a triorthogonal code is needed for a universal transversal gate set.

A basis of the orthogonal complement of $B_{\textrm{doubled}}$ is given by the rows of the matrix
\begin{equation}
\label{eq:doubledcomplement}
B_{\textrm{doubled}}^\bot = 
\begin{bmatrix}
B_{\textrm{sd},(0)} & B_{\textrm{sd},(0)} & 0_{\textrm{tri}} \\
0_{\textrm{sd}} & 1_{\textrm{sd}} & B_{\textrm{tri},(1)} \\
0_{\textrm{sd}} & 0_{\textrm{sd}} & B_{\textrm{tri},(0)} \\
B_{\textrm{sd},(0)} & 0_{\textrm{sd}} & 0_{\textrm{tri}}\\
E_{\textrm{sd}} & E_{\textrm{sd}} & 0_{\textrm{tri}}\\
0_{\textrm{sd}} & 0_{\textrm{sd}} & C_{\textrm{tri}}
\end{bmatrix} ,
\end{equation}
with $B_{\textrm{sd},(0)}$ as in Eq.~\ref{eq:selfdual}, $E_{\textrm{sd}}$ as in the discussion immediately following, $B_{\textrm{tri},(0)}$ and $B_{\textrm{sd},(0)}$ as in Eq.~\ref{eq:smalltriorthogonal}, and $C_{\textrm{tri}}$ as in the discussion immediately following. This particular construction explicitly separates the orthogonal complement, and the correspondingly constructed $Z$ stabilizers, into those that have a corresponding $X$ stabilizer in the doubled code and those that do not. A representation that mixes up these two types of stabilizers which can give a set of generators with lower weights is given similarly by the matrix
\begin{equation}
\label{eq:doubledcomplementalt}
B_{\textrm{doubled}}^\bot = 
\begin{bmatrix}
M_{\textrm{sd}} & M_{\textrm{sd}} & 0_{\textrm{tri}} \\
B_{\textrm{sd},(0)} & 0_{\textrm{sd}} & 0_{\textrm{tri}}\\
0_{\textrm{sd}} & B_{\textrm{sd},(1)} & L_{\textrm{tri},\textrm{Z}} \\
0_{\textrm{sd}} & 0_{\textrm{sd}} & B_{\textrm{tri}}^\bot
\end{bmatrix} ,
\end{equation}
where $M_{\textrm{sd},(0)}$ is any matrix which spans the same subspace as the matrices $B_{\textrm{sd},(0)}$ and $E_{\textrm{sd}}$ (which is the full even-weight subspace of $Z_{2}^{n}$, and can be generated using only rows of weight 2), $B_{\textrm{tri}}^\bot$ is any representation of the orthogonal complement of $B_{\textrm{tri}}$, and $L_{\textrm{tri},\textrm{Z}}$ is a representation of the logical Pauli $Z$ for the triorthogonal code. Note that in this form, the stabilizer correlating a self-dual block with the triorthogonal block, the one with the form $(0_{\textrm{sd}} , B_{\textrm{sd},(1)} , L_{\textrm{tri},\textrm{Z}})$, uses any representation of the logical of the self-dual code, not merely the fully transversal form, and also uses any representation of the logical $Z$ of the triorthogonal code, which will have shorter weight representations than $B_{\textrm{tri},(1)}$ due to the extra $Z$ stabilizers of the triorthogonal code.

The distance of this new code will now be addressed. From the starting representation of the logical Pauli operators, $(B_{\textrm{sd},(1)} , B_{\textrm{tri},(1)} , B_{\textrm{sd},(1)})$, from Eq.~\ref{eq:doubledcode}, the logical Pauli $Z$ can be shortened using any of the stabilizers constructed from Eq.~\ref{eq:doubledcomplement}.
A self-dual code encoding one logical qubit with $k$ $X$ stabilizers and $k$ $Z$ stabilizers has $n=2k+1$ qubits. Since the rows of $B_{\textrm{sd},(0)}$, $B_{\textrm{sd},(1)}$, and $E_{\textrm{sd}}$ span the $2k+1$-D vector space and $B_{\textrm{sd},(0)}$ and $E_{\textrm{sd}}$ have even weight rows, the rows of $B_{\textrm{sd},(0)}$ and $E_{\textrm{sd}}$ span the even-weight subspace of the $2k+1$-D vector space. Thus, the stabilizers of the form $(B_{\textrm{sd},(0)} , B_{\textrm{sd},(0)}, 0_{\textrm{tri}})$ and $(E_{\textrm{sd}} , E_{\textrm{sd}}, 0_{\textrm{tri}})$ can be used to reduce the first and second block of the logical $Z$ to only operating on one qubit each.
Stabilizers from the set $(0_{\textrm{sd}} , 0_{\textrm{sd}} , B_{\textrm{tri},(0)})$ and $(0_{\textrm{sd}} , 0_{\textrm{sd}} , C_{\textrm{tri}})$ can be used to make the third block of the logical $Z$ act on only $d_{\textrm{tri}}$ qubits, where $d_{\textrm{tri}}$ is the distance of the triorthogonal code constructed from $B_{\textrm{tri}}$. So the distance of this doubled code is at most $d_{\textrm{tri}} + 2$.

On the other hand, starting again from the original representation of the logical Pauli operators, $(B_{\textrm{sd},(1)} , B_{\textrm{sd},(1)} , B_{\textrm{tri},(1)}) $, using the stabilizers from the set of $(0_{\textrm{sd}}, 1_{\textrm{sd}} , B_{\textrm{tri},(1)})$ and $(B_{\textrm{sd},(0)} , B_{\textrm{sd},(0)} , 0_{\textrm{tri}})$, one can also reduce the logical $Z$ to the form $(1_{\textrm{sd}}, 0_{\textrm{sd}}, 0_{\textrm{tri}})$. From here, the stabilizers from the set $(B_{\textrm{sd},(0)} , 0_{\textrm{sd}} , 0_{\textrm{tri}})$ can be used to reduce the logical $Z$ down to acting on only $d_{\textrm{sd}}$ qubits, with $d_{\textrm{sd}}$ being the distance of the original self-dual code associated with the matrix $B_{\textrm{sd}}$. The distance $d_{\textrm{doubled}}$ of the doubled code associated with the triorthogonal matrix $B_{\textrm{doubled}}$ is thus
\begin{equation}
\label{eq:distance}
d_{\textrm{doubled}} = \min{(d_{\textrm{sd}}, d_{\textrm{tri}}+2)}.
\end{equation}
From this, it is clear why the general procedure needs a triorthogonal code instead of a single ancilla bit like the Reed-Muller code construction. With only the single ancilla bit (which can be seen as a trivial case of a distance 1 triorthogonal code), the resulting code can only ever be up to distance 3. Since the Steane code is also distance 3, that is perfectly fine for that code conversion technique.

\subsection{Converting to and from the doubled code}
\label{sec:convert}
One fault-tolerant code conversion between the self-dual code and the triorthogonal doubled code works as in the Steane and Reed-Muller conversion procedure~\cite{steane1996error}.  For going from the self-dual code to the doubled code:
\begin{enumerate}
\item Extend the self-dual code with entangled ancillas in the state $(\ket{0} \ket{0} + \ket{1} \ket{1})/\sqrt{2}$, with one of the logical qubits using the self-dual code and one using the smaller triorthogonal code
\item Measure the stabilizers of the doubled code
\item Perform error correction using only the syndromes of the stabilizers that were shared between the two forms
\item Fix the sign of the extra stabilizers  [the stabilizers of the form $(E_{\textrm{sd}} , E_{\textrm{sd}} , 0_{\textrm{tri}})$] using their measured syndromes
\end{enumerate}
For going from the doubled code to the self-dual code:
\begin{enumerate}
\item Measure the stabilizers of the extended form of the self-dual code with the entangled ancilla of the smaller triorthogonal code and the second self-dual code
\item Perform error correction using only the syndromes of the stabilizers that were shared between the two forms
\item Fix the sign of the extra stabilizers (the $X$ stabilizers of the form $(B_{\textrm{sd},(0)}, 0, 0)$ using their measured syndromes
\item Discard the additional ancilla
\end{enumerate}
The stabilizers from $B_{\textrm{doubled},(0)}$ are shared among both the doubled code and the ancilla-extended self-dual code. These shared stabilizers suffice to correct up to $k$ errors for $2k+1 = d_{\textrm{doubled}}$, which makes the procedure fault-tolerant\footnote{There are also $Z$ stabilizers of the form $(B_{\textrm{sd},(0)}, 0, 0)$ which are shared among both forms, but they are not needed to correct errors of weight $k$ or less; their syndromes can simply be fixed with Pauli corrections that commute with the other stabilizers and the logicals.}.

\begin{figure}
\begin{center}
\includegraphics{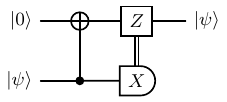} \includegraphics{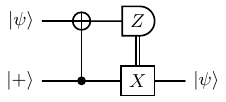}
\end{center}
\caption{Gadgets that move a qubit into an ancilla, which are useful for code conversion. }
\label{fig:swapgadget}
\end{figure}

\begin{figure}
\begin{center}
\includegraphics{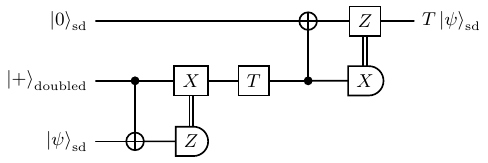}
\end{center}
\caption{Example of a circuit to perform a logical $T$ gate on a qubit encoded in the self-dual code, via code conversion to a doubled code and back. Every operation is transversal in this procedure, with the $CNOT$ gates involving the doubled code using only the first self-dual block of the code.}
\label{fig:Tgadget}
\end{figure}

There is also another procedure for converting between the two codes. The extra $Z$ stabilizers of the doubled code allow for a transversal $CNOT$ between a logical qubit encoded using the doubled code as a control, using only the qubits in the first block, and a logical qubit encoded using the original self-dual code as the target\footnote{This transversal $CNOT$ works also for the Reed-Muller code and Steane code; this fact does not appear to be noted in the literature.}.  Although this logical $CNOT$ only works in one direction, using the gadgets shown in Fig.~\ref{fig:swapgadget}, this can be used to convert between the two codes using encoded ancilla. To convert from the doubled code to the self-dual code, a $\ket{0}$ ancilla encoded in the self-dual code is consumed. To convert from the self-dual code to the doubled code, a $\ket{+}$ ancilla encoded in the doubled code is consumed. These are the simple sort of ancilla that would already be used with Steane-style syndrome extraction. The sort of circuit to implement a $T$ gate via this style of code conversion is shown in Fig.~\ref{fig:Tgadget}

With the appropriate choice of a smaller triorthogonal code, this construction thus gives a procedure to fault-tolerantly convert between a self-dual CSS code (which has transversal implementations for $H$, $S$ and $CNOT$) and a constructed triorthogonal code which supports a transversal implementation of the logical $T$ gate, thus giving access to a fault-tolerant universal gate set.

\section{Minimum distance decoding}
\label{sec:syndrome}
To make full use of this code-doubling construction, it is necessary to know how to correct errors. There is much that can be said about syndrome measurement itself; see e.g. Reference~\cite{PhysRevA.97.032331} and Reference~\cite{paetznick2012golay} which analyzed the preparation of ancillas for Steane's method for extracting syndromes for large codes, with specific application to the Golay code. General ancilla preparation techniques will be useful for the doubled code, but there is not a straightforward way to directly convert ancilla preparation procedures for the self-dual code and triorthogonal code into ancilla preparation procedures for the doubled code. For example, to prepare noisy $\ket{0}_{\textrm{doubled}}$ states with Steane's latin rectangle methods\cite{steane2002codewords}, the Gaussian elimination of $B_{\textrm{doubled},(0)}$ does not simply reduce to Gaussian elimination on $B_{\textrm{sd},(0)}$ and $B_{\textrm{tri},(0)}$ due to the large $X$ stabilizer with the form $(0_{\textrm{sd}} , 1_{\textrm{sd}} , B_{\textrm{tri},(1)})$.

Assuming the successful extraction of the error syndromes, the next question, which shall be addressed in this section, is then how to perform the error correction for a given syndrome in the code-doubling construction. Specifically, shortest error decoding using the shared $X$ and $Z$ stabilizers from only $B_{\textrm{doubled},(0)}$ will be addressed. Appendix B of reference~\cite{PhysRevA.93.052332} includes a similar discussion, although there is a special case, to be discussed later, which requires a more careful treatment for proper decoding.

There are three kinds of stabilizers in $B_{\textrm{doubled},(0)}$:
\begin{enumerate}
\item Stabilizers acting only on the two blocks of the self-dual code,
\item Stabilizers acting only on the block of the smaller triorthogonal code,
\item Stabilizer correlating one of the self-dual blocks with the triorthogonal block.
\end{enumerate}
For the first two kinds of stabilizers, the correction procedure reduces to the shortest error correction procedure for the two kinds of codes used in the construction. The syndromes for the first kind of stabilizer can be corrected by applying the error correction procedure for the self-dual code to the first block only, and the syndromes for the second kind of stabilizer can be corrected using the error correction procedure for the triorthogonal code. The third kind of stabilizer, the one of the form $(0_{\textrm{sd}} , 1_{\textrm{sd}} , B_{\textrm{tri},(1)})$, however, provides a slight complication. It is possible that the procedure for correcting the first two kinds of stabilizers will already leave the syndrome of the third kind of stabilizer with the correct $+1$ value, based on how many of the qubit corrections for the triorthogonal block overlap with $B_{\textrm{tri},(1)}$. However, if the existing corrections will not fix the last stabilizer, then an additional correction is needed. As long as at least one qubit of the first self-dual block is going to be corrected, then the solution is simple: move one of the qubit corrections from the first self-dual block to the second self-dual block. 

The special case that needs to be carefully considered is when there are no corrections required for the self-dual code blocks and yet the $(0_{\textrm{sd}} , 1_{\textrm{sd}} , B_{\textrm{tri},(1)})$ stabilizer needs additional correction. There are two distinct ways to change this stabilizer while leaving the other stabilizers unchanged: either apply a Pauli correction to one qubit in the first self-dual code block and to the same qubit in the second self-dual code block, or apply a logical Pauli operation to the smaller triorthogonal code block. The two options have opposite effect on the logical operator of the doubled code, so care needs to be taken to choose whichever is shorter\footnote{Reference~\cite{PhysRevA.93.052332} suggests always applying single qubit corrections to both self-dual blocks. The quantum Reed-Muller code, which can only correct 1 $Z$ error, can be seen as a special case of this construction with a trivial triorthogonal block of one single qubit. Always applying a correction of weight 2 in this case would not decode the shortest error. This shows the need to distinguish which correction to apply.}. To accurately determine this, one not only needs to know the weight of the shortest representation of the error $E$ for any given syndrome, but also the weight of the shortest representation of $E+L$, with $L$ a logical Pauli operator, for the triorthogonal code. Then whether the minimum weight of $E+L$ is smaller or larger than two plus the smallest weight of $E$ determines which correction to make\footnote{For some situations, the answer will be obvious based only on the smallest weight of $E$ and the distance of the doubled code. However, when some subset of errors larger than $\frac{d-1}{2}$ are also correctable, then explicit information about the smallest weight of $E+L$ might be necessary sometimes.}.

\section{Application to the quantum Golay code}
\label{sec:golay}
\subsection{Constructing the $[[95,1,7]]$ triorthogonal code}
\label{sec:golay95}
Using the same notation as before for representing self-dual CSS codes with a matrix, the $[[23, 1, 7]]$ quantum Golay code is represented as 
\begin{widetext}
\begin{equation}
\label{eq:golay}
B_{\textrm{G}}= 
\begin{bmatrix}
 1 & 1 & 1 & 1 & 1 & 1 & 1 & 1 & 1 & 1 & 1 & 1 & 1 & 1 & 1 & 1 & 1 & 1 & 1 & 1 & 1 & 1 & 1 \\ \hline
 1 & 1 & 1 & 1 & 0 & 0 & 1 & 0 & 0 & 1 & 0 & 1 & 0 & 0 & 0 & 0 & 0 & 0 & 0 & 0 & 0 & 0 & 1 \\
 0 & 0 & 0 & 1 & 0 & 1 & 1 & 0 & 1 & 1 & 1 & 1 & 0 & 0 & 0 & 0 & 0 & 0 & 0 & 0 & 0 & 1 & 0 \\
 0 & 0 & 1 & 0 & 1 & 1 & 0 & 1 & 1 & 1 & 1 & 0 & 0 & 0 & 0 & 0 & 0 & 0 & 0 & 0 & 1 & 0 & 0 \\
 0 & 1 & 0 & 1 & 1 & 0 & 1 & 1 & 1 & 1 & 0 & 0 & 0 & 0 & 0 & 0 & 0 & 0 & 0 & 1 & 0 & 0 & 0 \\
 1 & 0 & 1 & 1 & 0 & 1 & 1 & 1 & 1 & 0 & 0 & 0 & 0 & 0 & 0 & 0 & 0 & 0 & 1 & 0 & 0 & 0 & 0 \\
 1 & 0 & 0 & 1 & 1 & 1 & 0 & 1 & 0 & 1 & 0 & 1 & 0 & 0 & 0 & 0 & 0 & 1 & 0 & 0 & 0 & 0 & 0 \\
 1 & 1 & 0 & 0 & 1 & 0 & 0 & 0 & 1 & 1 & 1 & 1 & 0 & 0 & 0 & 0 & 1 & 0 & 0 & 0 & 0 & 0 & 0 \\
 0 & 1 & 1 & 0 & 0 & 0 & 1 & 1 & 1 & 0 & 1 & 1 & 0 & 0 & 0 & 1 & 0 & 0 & 0 & 0 & 0 & 0 & 0 \\
 1 & 1 & 0 & 0 & 0 & 1 & 1 & 1 & 0 & 1 & 1 & 0 & 0 & 0 & 1 & 0 & 0 & 0 & 0 & 0 & 0 & 0 & 0 \\
 0 & 1 & 1 & 1 & 1 & 1 & 0 & 0 & 1 & 0 & 0 & 1 & 0 & 1 & 0 & 0 & 0 & 0 & 0 & 0 & 0 & 0 & 0 \\
 1 & 1 & 1 & 1 & 1 & 0 & 0 & 1 & 0 & 0 & 1 & 0 & 1 & 0 & 0 & 0 & 0 & 0 & 0 & 0 & 0 & 0 & 0
\end{bmatrix} 
=\begin{bmatrix}B_{\textrm{G},(1)} \\ \hline B_{\textrm{G},(0)} \end{bmatrix} .
\end{equation}
\end{widetext}
One possible choice for an appropriate even-weight matrix with which to form a complete basis together with the rows of $B_{\textrm{G}}$, which can be obtained from the product of 11 selected pairs of rows of $B_{\textrm{G},(0)}$, is
\begin{widetext}
\begin{equation}
\label{eq:golayE}
E_{\textrm{G}} = \begin{bmatrix}
 0 & 0 & 0 & 1 & 0 & 0 & 1 & 0 & 0 & 1 & 0 & 1 & 0 & 0 & 0 & 0 & 0 & 0 & 0 & 0 & 0 & 0 & 0 \\
 0 & 0 & 0 & 0 & 0 & 1 & 0 & 0 & 1 & 1 & 1 & 0 & 0 & 0 & 0 & 0 & 0 & 0 & 0 & 0 & 0 & 0 & 0 \\
 0 & 0 & 0 & 0 & 1 & 0 & 0 & 1 & 1 & 1 & 0 & 0 & 0 & 0 & 0 & 0 & 0 & 0 & 0 & 0 & 0 & 0 & 0 \\
 0 & 0 & 0 & 1 & 0 & 0 & 1 & 1 & 1 & 0 & 0 & 0 & 0 & 0 & 0 & 0 & 0 & 0 & 0 & 0 & 0 & 0 & 0 \\
 1 & 0 & 0 & 1 & 0 & 1 & 0 & 1 & 0 & 0 & 0 & 0 & 0 & 0 & 0 & 0 & 0 & 0 & 0 & 0 & 0 & 0 & 0 \\
 1 & 0 & 0 & 0 & 1 & 0 & 0 & 0 & 0 & 1 & 0 & 1 & 0 & 0 & 0 & 0 & 0 & 0 & 0 & 0 & 0 & 0 & 0 \\
 0 & 1 & 0 & 0 & 0 & 0 & 0 & 0 & 1 & 0 & 1 & 1 & 0 & 0 & 0 & 0 & 0 & 0 & 0 & 0 & 0 & 0 & 0 \\
 0 & 1 & 0 & 0 & 0 & 0 & 1 & 1 & 0 & 0 & 1 & 0 & 0 & 0 & 0 & 0 & 0 & 0 & 0 & 0 & 0 & 0 & 0 \\
 0 & 1 & 0 & 0 & 0 & 1 & 0 & 0 & 0 & 0 & 0 & 0 & 0 & 0 & 0 & 0 & 0 & 0 & 0 & 0 & 0 & 0 & 0 \\
 0 & 1 & 1 & 1 & 1 & 0 & 0 & 0 & 0 & 0 & 0 & 0 & 0 & 0 & 0 & 0 & 0 & 0 & 0 & 0 & 0 & 0 & 0 \\
 0 & 1 & 0 & 1 & 0 & 0 & 1 & 0 & 0 & 1 & 0 & 0 & 0 & 0 & 0 & 0 & 0 & 0 & 0 & 0 & 0 & 0 & 0
\end{bmatrix}.
\end{equation}
\end{widetext}

To construct a distance 7 triorthogonal code with this code-doubling scheme using the quantum Golay code, a distance 5 triorthogonal code is the remaining necessary ingredient. There is a $[[49,1,5]]$ code known in the literature which is suitable\cite{bravyi2012magic}.  One could imagine, however, using this same construction with a distance 5 self-dual CSS code and a distance 3 triorthogonal code in order to construct a distance 5 code. In fact, the $[[49,1,5]]$ code in the literature is already equivalent to such a code constructed from a $[[17,1,5]]$ self-dual CSS code (which is, in fact, a color code), and the $[[15,1,3]]$ quantum Reed-Muller code. The matrix representation of the $[[49,1,5]]$ triorthogonal code can be written in block form as
\begin{equation}
\label{eq:tri49}
B_{49} =
\begin{bmatrix}
B_{17,(1)} & B_{17,(1)} & B_{\textrm{RM},(1)} \\
\hline
B_{17,(0)} & B_{17,(0)} & 0_{\textrm{RM}} \\
0_{17} & 1_{17} & B_{\textrm{RM},(1)}\\
0_{17} & 0_{17} & B_{\textrm{RM},(0)}
\end{bmatrix},
\end{equation}
with the submatrices coming from the $[[17,1,5]]$ self-dual color code,
%\begin{widetext}
\begin{eqnarray}
\label{eq:sd17}
B_{17} &=& 
\begin{bmatrix}
 1 & 1 & 1 & 1 & 1 & 1 & 1 & 1 & 1 & 1 & 1 & 1 & 1 & 1 & 1 & 1 & 1 \\ \hline
 1 & 1 & 0 & 0 & 0 & 1 & 1 & 0 & 0 & 0 & 0 & 0 & 0 & 0 & 0 & 0 & 0 \\
 0 & 0 & 0 & 0 & 0 & 1 & 1 & 0 & 0 & 1 & 1 & 0 & 0 & 0 & 0 & 0 & 0 \\
 0 & 0 & 0 & 0 & 0 & 0 & 0 & 0 & 0 & 1 & 1 & 0 & 0 & 1 & 0 & 1 & 0 \\
 0 & 0 & 0 & 0 & 0 & 0 & 0 & 0 & 0 & 0 & 0 & 0 & 0 & 1 & 1 & 1 & 1 \\
 0 & 1 & 1 & 0 & 0 & 0 & 1 & 1 & 0 & 0 & 1 & 1 & 0 & 1 & 1 & 0 & 0 \\
 0 & 0 & 1 & 1 & 0 & 0 & 0 & 1 & 1 & 0 & 0 & 0 & 0 & 0 & 0 & 0 & 0 \\
 0 & 0 & 0 & 0 & 0 & 0 & 0 & 1 & 1 & 0 & 0 & 1 & 1 & 0 & 0 & 0 & 0 \\
 0 & 0 & 0 & 1 & 1 & 0 & 0 & 0 & 1 & 0 & 0 & 0 & 1 & 0 & 0 & 0 & 0
\end{bmatrix} \nonumber \\
&=&
\begin{bmatrix}
B_{17,(1)} \\ \hline
B_{17,(0)}
\end{bmatrix},
\end{eqnarray}
%\end{widetext}
and the $[[15,1,3]]$ triorthogonal Reed-Muller code,
%\begin{widetext}
\begin{eqnarray}
\label{eq:RM}
B_{\textrm{RM}}
&=& 
\begin{bmatrix}
 1 & 1 & 1 & 1 & 1 & 1 & 1 & 1 & 1 & 1 & 1 & 1 & 1 & 1 & 1 \\ \hline
 1 & 0 & 1 & 0 & 1 & 0 & 1 & 0 & 1 & 0 & 1 & 0 & 1 & 0 & 1 \\
 0 & 1 & 1 & 0 & 0 & 1 & 1 & 0 & 0 & 1 & 1 & 0 & 0 & 1 & 1 \\
 0 & 0 & 0 & 1 & 1 & 1 & 1 & 0 & 0 & 0 & 0 & 1 & 1 & 1 & 1 \\
 0 & 0 & 0 & 0 & 0 & 0 & 0 & 1 & 1 & 1 & 1 & 1 & 1 & 1 & 1 
\end{bmatrix} \nonumber \\
&=&
\begin{bmatrix}
B_{\textrm{RM},(1)} \\ \hline
B_{\textrm{RM},(0)}
\end{bmatrix}.
\end{eqnarray}
%\end{widetext}
For the convenient construction of a basis for the orthogonal complement of the $[[49,1,5]]$ triorthogonal matrix as in Eq.~\ref{eq:doubledcomplement}, two more matrices are required: an even-weight matrix $E_{17}$ whose rows, together with the rows of $B_{17}$, form a complete basis of $Z_{2}^{17}$, and a matrix $C_{\textrm{RM}}$ whose rows, together with the rows of $B_{\textrm{RM},(0)}$ span the orthogonal complement of $B_{\textrm{RM}}$. One such choice for these matrices is
%\begin{widetext}
\begin{equation}
\label{eq:sd17extra}
E_{17} = 
\begin{bmatrix}
 0 & 0 & 0 & 0 & 0 & 1 & 1 & 0 & 0 & 0 & 0 & 0 & 0 & 0 & 0 & 0 & 0 \\
 0 & 1 & 0 & 0 & 0 & 0 & 1 & 0 & 0 & 0 & 0 & 0 & 0 & 0 & 0 & 0 & 0 \\
 0 & 0 & 0 & 0 & 0 & 0 & 1 & 0 & 0 & 0 & 1 & 0 & 0 & 0 & 0 & 0 & 0 \\
 0 & 0 & 0 & 0 & 0 & 0 & 0 & 0 & 0 & 0 & 1 & 0 & 0 & 1 & 0 & 0 & 0 \\
 0 & 0 & 0 & 0 & 0 & 0 & 0 & 0 & 0 & 0 & 0 & 0 & 0 & 1 & 1 & 0 & 0 \\
 0 & 0 & 1 & 0 & 0 & 0 & 0 & 1 & 0 & 0 & 0 & 0 & 0 & 0 & 0 & 0 & 0 \\
 0 & 0 & 0 & 0 & 0 & 0 & 0 & 1 & 0 & 0 & 0 & 1 & 0 & 0 & 0 & 0 & 0 \\
 0 & 0 & 0 & 0 & 0 & 0 & 0 & 1 & 1 & 0 & 0 & 0 & 0 & 0 & 0 & 0 & 0
\end{bmatrix}
\end{equation}
%\end{widetext}
and
%\begin{widetext}
\begin{equation}
\label{eq:RMextra}
C_{\textrm{RM}} = 
\begin{bmatrix}
 0 & 0 & 1 & 0 & 0 & 0 & 1 & 0 & 0 & 0 & 1 & 0 & 0 & 0 & 1 \\
 0 & 0 & 0 & 0 & 1 & 0 & 1 & 0 & 0 & 0 & 0 & 0 & 1 & 0 & 1 \\
 0 & 0 & 0 & 0 & 0 & 0 & 0 & 0 & 1 & 0 & 1 & 0 & 1 & 0 & 1 \\
 0 & 0 & 0 & 0 & 0 & 1 & 1 & 0 & 0 & 0 & 0 & 0 & 0 & 1 & 1 \\
 0 & 0 & 0 & 0 & 0 & 0 & 0 & 0 & 0 & 1 & 1 & 0 & 0 & 1 & 1 \\
 0 & 0 & 0 & 0 & 0 & 0 & 0 & 0 & 0 & 0 & 0 & 1 & 1 & 1 & 1
\end{bmatrix},
\end{equation}
%\end{widetext}
both of which are constructed similarly to $E_{\textrm{G}}$ from Eq.~\ref{eq:golayE}. Using these matrices, the construction from Eq.~\ref{eq:doubledcomplement} gives
\begin{equation}
\label{eq:49complement}
B_{49}^\bot = 
\begin{bmatrix}
B_{17,(0)} & B_{17,(0)} & 0_{\textrm{RM}}  \\
0_{17} & 1_{17} & B_{\textrm{RM},(1)} \\
0_{17} & 0_{17} & B_{\textrm{RM},(0)} \\
B_{17,(0)} & 0_{17} & 0_{\textrm{RM}}\\
E_{17} & E_{17} & 0_{\textrm{RM}}\\
0_{17} & 0_{17} & C_{\textrm{RM}}
\end{bmatrix} = 
\begin{bmatrix}
B_{49,(0)}\\
C_{49}
\end{bmatrix}
\end{equation}
as a basis for the orthogonal complement, and thus the $Z$ stabilizers for the 49 qubit triorthogonal code, with $C_{49}$ being the additional set of $Z$ stabilizers without corresponding $X$ stabilizers.

Using the code-doubling procedure with the $[[23,1,7]]$ quantum Golay code and the $[[49,1,5]]$ triorthogonal code, the construction yields a $[[95,1,7]]$ triorthogonal code which can be converted back and forth to the quantum Golay code fault-tolerantly. The matrix representation $B_{95}$ of the triorthogonal code in block form is
%\begin{widetext}
\begin{equation}
\label{eq:tri95}
B_{95} =
\begin{bmatrix}
B_{\textrm{G},(1)} & B_{\textrm{G},(1)} & B_{17,(1)} & B_{17,(1)} & B_{\textrm{RM},(1)} \\
\hline
B_{\textrm{G},(0)} & B_{\textrm{G},(0)} & 0_{17} & 0_{17} & 0_{\textrm{RM}} \\
0_{\textrm{G}} & 1_{\textrm{G}} & B_{17,(1)} & B_{17,(1)} & B_{\textrm{RM},(1)} \\
0_{\textrm{G}} & 0_{\textrm{G}} & B_{17,(0)} & B_{17,(0)} & 0_{\textrm{RM}} \\
0_{\textrm{G}} & 0_{\textrm{G}} & 0_{17} & 1_{17} & B_{\textrm{RM},(1)}\\
0_{\textrm{G}} & 0_{\textrm{G}} & 0_{17} & 0_{17} & B_{\textrm{RM},(0)}
\end{bmatrix},
\end{equation}
%\end{widetext}
and a basis for the orthogonal complement can be given by the rows of the matrix
%\begin{widetext}
\begin{eqnarray}
\label{eq:95complement}
B_{95}^\bot &=& 
\begin{bmatrix}
B_{\textrm{G},(0)} & B_{\textrm{G},(0)} & 0_{17} & 0_{17} & 0_{\textrm{RM}} \\
0_{\textrm{G}} & 1_{\textrm{G}} & B_{17,(1)} & B_{17,(1)} & B_{\textrm{RM},(1)} \\
0_{\textrm{G}} & 0_{\textrm{G}} & B_{17,(0)} & B_{17,(0)} & 0_{\textrm{RM}} \\
0_{\textrm{G}} & 0_{\textrm{G}} & 0_{17} & 1_{17} & B_{\textrm{RM},(1)}\\
0_{\textrm{G}} & 0_{\textrm{G}} & 0_{17} & 0_{17} & B_{\textrm{RM},(0)}\\
0_{\textrm{G}} & B_{\textrm{G},(0)} & 0_{17} & 0_{17} & 0_{\textrm{RM}} \\
E_{\textrm{G}} & E_{\textrm{G}} & 0_{17} & 0_{17} & 0_{\textrm{RM}} \\
0_{\textrm{G}} & 0_{\textrm{G}} & B_{17,(0)} & 0_{17} & 0_{\textrm{RM}}\\
0_{\textrm{G}} & 0_{\textrm{G}} & E_{17} & E_{17} & 0_{\textrm{RM}}\\
0_{\textrm{G}} & 0_{\textrm{G}} & 0_{17} & 0_{17} & C_{\textrm{RM}}
\end{bmatrix} \nonumber \\
&=& 
\begin{bmatrix}
B_{95,(0)}\\
C_{95}
\end{bmatrix},
\end{eqnarray}
%\end{widetext}
with $C_{95}$ giving the additional set of $Z$ stabilizers for the doubled code.
\subsection{Correcting errors}
\label{sec:golay95errors}
The three component codes whose error correction must be known to correct errors for the $[[95,1,7]]$ triorthogonal code are the $[[23,1,7]]$ quantum Golay code, the $[[17,1,5]]$ color code, and the $[[15,1,3]]$ quantum Reed-Muller code. Each of these reduces to error correction on a corresponding classical code. These will only be briefly summarized here. For the $[[23,1,7]]$ quantum Golay code, all errors on up to 3 qubits are correctable, and also {\em only} errors with a minimum weight of 3 or lower are correctable. For the $[[17,1,5]]$ color code, all errors on up to 2 qubits are correctable, but there is also a subset of errors of weight 3 that are correctable. And for the $[[15,1,3]]$ quantum Reed-Muller code, only errors on a single qubit can be corrected using only the shared $X$ and $Z$ stabilizers. 

\begin{table}
\centering
\begin{tabular}{c|c|c||c|c|c}
color & Reed-Muller & parity & corrections & $w(E)$ & $w(E+L)$\\
\hline
 0 & 0 & no  &      & 0 & 5 \\
 0 & 0 & yes & $+2$ & 2 & 3 \\
 0 & 1 & no  &      & 1 & 4 \\
 0 & 1 & yes & $+L$ & 2 & 3 \\
 1 & 0 & no  &      & 1 & 4 \\
 1 & 0 & yes & move & 1 & 4 \\
 1 & 1 & no  &      & 2 & 3 \\
 1 & 1 & yes & move & 2 & 3 \\
 2 & 0 & no  &      & 2 & 3 \\
 2 & 0 & yes & move & 2 & 3 \\
 2 & 1 & no  &      & 3 & 4 \\
 2 & 1 & yes & move & 3 & 4 \\
 3 & 0 & no  &      & 3 & 4 \\
 3 & 0 & yes & move & 3 & 4 \\
 3 & 1 & no  &      & 4 & 5 \\
 3 & 1 & yes & move & 4 & 5 
\end{tabular}

\caption{Additional error correction information for the $[[49,1,5]]$ triorthogonal code. The first two columns indicate the number of errors indicated by the naive interpretation of the syndromes for the color code blocks (`color') and quantum Reed-Muller code block (`Reed-Muller'). The column `parity' indicates whether the stabilizer correlating the two component code blocks requires additional corrections. The `corrections' column indicates the extra correction to apply in  that case: $+L$ means applying an additional logical Pauli to the inner Reed-Muller block, $+2$ means applying a single qubit Pauli correction to the same location in both color code blocks, and `move' means moving one qubit correction from the first color code block onto the second color code block. The last two columns are the minimum weights of the error $E$ and the error plus logical Pauli $E+L$.}
\label{table:errorprocedure49}
\end{table}

As mentioned in Section~\ref{sec:syndrome}, the error decoding procedure for the $[[95,1,7]]$ doubled code reduces to the slightly modified decoding procedure for the $[[23,1,7]]$ quantum Golay code and the $[[49,1,5]]$ triorthogonal code, and the decoding procedure for the $[[49,1,5]]$ code itself reduces to the slightly modified procedure for the $[[17,1,5]]$ color code and the $[[15,1,3]]$ quantum Reed-Muller code. The only special cases to consider are when the self-dual block has no errors and when the stabilizer correlating the self-dual block and component triorthogonal block needs additional correction. The relevant information for the error correction of the $[[49,1,5]]$ code is described in Table~\ref{table:errorprocedure49}. Note that even some subset of errors of up to weight 4 are correctable for this code.

\begin{table}
\centering
\begin{tabular}{c|c|c||c}
Golay & 49 qubit & parity & corrections\\
\hline
 0 & 0 & no  &      \\
 0 & 0 & yes & $+2$ \\
 0 & 1 & no  &      \\
 0 & 1 & yes & $+2$ \\
 0 & 2 & no  &      \\
 0 & 2 & yes & $+L$ \\
 0 & 3 & no  &      \\
 0 & 3 & yes & $+L$ \\
 0 & 4 & no  &      \\
 0 & 4 & yes & $+L$ \\
 $\geq$ 1 & any & no & \\
 $\geq$ 1 & any & yes & move \\
\end{tabular}

\caption{Additional error correction information for the $[[95,1,7]]$ triorthogonal code. The first two columns indicate the number of errors indicated by the naive interpretation of the syndromes for the Golay code blocks (`Golay') and 49-qubit triorthogonal code block (`49 qubit'). The column `parity' indicates whether the stabilizer correlating the two component code blocks requires additional corrections. The `corrections' column indicates the extra correction to apply in  that case: $+L$ means applying an additional logical Pauli to the inner 49-qubit triorthogonal block, $+2$ means applying a single qubit Pauli correction to the same location in both Golay code blocks, and `move' means moving one qubit correction from the first Golay code block onto the second Golay code block.}
\label{table:errorprocedure95}
\end{table}
Using the information on the weights from Table~\ref{table:errorprocedure49} for the $[[49,1,5]]$ code, the additional error correction step for the $[[95,1,7]]$ doubled code can be determined, and is summarized in Table~\ref{table:errorprocedure95}. If the stabilizer correlating the $[[49,1,5]]$ triorthogonal block with one of the $[[23,1,7]]$ Golay blocks requires additional correction and there are no detected errors on the Golay blocks, then the additional correction will either be
\begin{enumerate}
\item applying an additional logical Pauli to the $[[49,1,5]]$ block, or
\item applying a single qubit Pauli correction to the same location in both Golay code blocks.
\end{enumerate}
Note that the weights $w(E+L)$ in Table~\ref{table:errorprocedure49} are uniquely determined by $w(E)$ for the $[[49,1,5]]$ code, so determining which of the two procedures is shorter can be determined solely from the weight of the error as determined by the $[[49,1,5]]$ syndromes regardless of their structure on the subblocks. If the weight of the error detected using only the $[[49,1,5]]$ syndromes is either 0 or 1, then option 2 is the correct option for the shortest error decoding. If the weight of the error is 2 or larger, then option 1 is the correct option\footnote{Performing the correction associated with option 1 using the actual shortest representation of the error and logical is not required but can be useful to minimize the number of gates applied.}. Note that while all errors of up to weight 3 are correctable for the $[[95,1,7]]$ code, since some errors of up to weight 4 are correctable on the $[[49,1,5]]$ block and errors of up to weight 3 are correctable on the $[[23,1,7]]$ blocks, even some subset of errors up to weight 7 are correctable by this code. 

\section{Using doubled codes for a fault-tolerant universal gate set}
\label{sec:disc}
\begin{figure}
\begin{centering}
\includegraphics{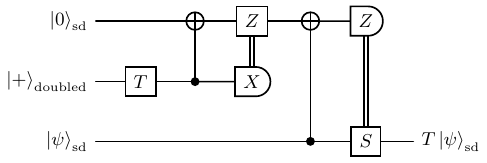}
\end{centering}
\caption{An example circuit for preparing high-quality encoded magic states with transversal $T$ gates using the doubled code, which are then converted to the self-dual code and consumed as in normal magic state procedures.}
\label{fig:magicgadget}
\end{figure}

The construction of this doubled code allows for a few different ways to implement fault-tolerant universal quantum computation. The obvious one is one of the code conversion procedures already described. Clifford gates can be applied to the $[[23,1,7]]$ quantum Golay code, and then the code can be converted to the $[[95,1,7]]$ code when $T$ gates are necessary, and the conversion can be done back and forth depending on what gates are necessary. One related possibility is to directly prepare encoded magic states using the transversal $T$ gate of the $[[95,1,7]]$ code, convert these magic states into the $[[23,1,7]]$ code, and use standard magic state methods, as shown in Fig.~\ref{fig:magicgadget}. This replaces the need for using magic state distillation to produce high-quality encoded magic states with code conversion. Note that the gadget in Fig.~\ref{fig:magicgadget} has the same gate count, circuit depth, and consumed ancilla as the conversion gadget in Fig.~\ref{fig:Tgadget}, but with the final classically conditioned gate being a Clifford gate instead of a Pauli gate.

Another possibility for fault-tolerance is to strictly use the $[[95,1,7]]$ triorthogonal code. Logical Pauli, $CNOT$, $S$, and $T$ gates can be implemented transversally with this code. Using the procedure of Ref.~\cite{paetznick2013universal}, the logical Hadamard can be implemented by transversal application of Hadamard followed by error correction. This gives a fault-tolerant universal gate set on the doubled code.

There is another interesting avenue to mention: the concatenation of two distinct codes with different transversal gate sets\cite{PhysRevLett.112.010505}. This has been considered using the $[[7,1,3]]$ Steane code and the $[[15,1,3]]$ quantum Reed-Muller code. The concatenated code is a $[[105,1,9]]$ code, but it does not have transversal implementations of some necessary gates. However, $H$ can be implemented non-transversally at the level of the $[[15,1,3]]$ code and transversally at the level of the $[[7,1,3]]$ code, while the reverse is possible for the $T$ gate. This preserves the ability to correct single-qubit errors after applying the $H$ and $T$ gates. The gates that are transversal for both codes ($CNOT$, $S$, and Pauli gates) have the full distance 9 protection, enabling correction of arbitrary errors on up to 4 qubits. The full protection for $CNOT$ gates in particular is one of the more appealing points, as the asymptotic threshold of the code in fact ends up limited by the pseudo-threshold for the $CNOT$ gates rather than $H$ or $T$\cite{PhysRevLett.117.010501}. One could perform the same sort of concatenation procedure with the $[[23,1,7]]$ and $[[95,1,7]]$ codes to create a $[[2185,1,49]]$, which can correct errors on up to 24 qubits for $CNOT$, $S$, and Pauli gates and up to 3 errors for $T$ and $H$ gates. Such a code, while very large, may be a plausible avenue for fault tolerant quantum computation in the future with larger devices.

The code doubling procedure can generally be applied to a large class of self-dual CSS codes. One disadvantage is that for large codes, the procedure yields large weight stabilizers. Although code doubling was presented in the context of code conversion between a self-dual CSS code and a triorthogonal code, the procedure also can be used as a means of constructing triorthogonal codes with larger distances. Triorthogonal codes are useful tools for magic state distillation, although the distillation procedures tend to benefit from codes with a high yield rather than a high distance\cite{bravyi2012magic}.

Due to the size of the doubled code, implementation of the above quantum error correcting schemes on hardware in the near-term would be difficult. The Steane and quantum Reed-Muller codes are both rather small at 7 and 15 qubits, respectively. The quantum Golay code, at 23 qubits, is bigger than both of those codes combined. On top of that, the doubled code presented requires more than 4 times the qubits of the quantum Golay code. Measuring the stabilizers of these larger codes also generally requires more resources, e.g. in the number of $CNOT$ gates needed to prepare ancilla for Steane-style syndrome measurement. In spite of this, the benefit over the Steane code and quantum Reed-Muller code is the larger code distance protecting against larger errors, potentially obtaining a higher error threshold and suppressing error rates more rapidly with concatenation. When larger devices become available, these larger codes may prove advantageous.  A detailed analysis of the application of the various methods for this doubled code scheme could be a subject for future work.

\subsection{Comparison with magic state distillation schemes}
\label{sec:distillation}
The scheme presented in Fig.~\ref{fig:magicgadget} allows for the preparation of magic states in the doubled code and their consumption in the original quantum Golay code, as with typical magic state schemes. It is thus of interest to compare this scheme using the doubled code scheme with magic state distillation for the Golay code. A detailed numerical study using Steane-style error correction would require optimizing ancilla preparation circuits for the doubled code similarly to Ref.~\cite{paetznick2012golay}. It is not clear whether there would be a similar scheme to the Golay code which can use only four noisy ancilla to produce a high quality ancilla. However, a rough analysis shows that the qubit cost of the scaling could be favorable for the code doubling scheme regardless.

For a physical error rate of $p$, the doubled code, like the Golay code, has a logical error rate of $\mathcal{O}(p^4)$ Magic state distillation using a $3k+8$-to-$k$ scheme has $\mathcal{O}(p^2)$ error suppression, and thus requires two rounds to achieve the same $\mathcal{O}(p^4)$ error rate. Thus in the limit of large $k$, this family of magic state distillation schemes require 9 noisy encoded magic states for every 1 high-fidelity encoded magic state. At this point, the qubit count for the just the 9 noisy encoded ancilla per high-fidelity encoded magic state produced, a total of $23 \times 9 = 207$ physical qubits, compares favorably with the 95 qubits for the doubled code. Considering specifically Steane-style ancilla preparation for the doubled code, assuming the pessimistic case that 12 noisy ancilla are required, the physical qubit cost increases to 1140 for the doubled code per encoded magic state generated. However, in the magic state distillation procedure, the analogous Steane-style preparation of ancilla for measuring $X$ and $Z$ stabilizers of the initial 9 magic states is required. With the optimized procedures from Ref.~\cite{paetznick2012golay}, each high-fidelity $\ket{0}$ and $\ket{+}$ can be produced from 4 noisy ancilla. This raises the requirements for magic state distillation with Steane-style error correction to 1863 physical qubits, just for the first round. The full magic state distillation procedure will also involve other encoded qubits which adds to the resource cost further. The final error correction in both procedures is done in the Golay code and increases the qubit cost by the same amount for each. With Shor-style error correction, the counting changes for both procedures, and becomes closer to the original qubit counting of 95 and 207.

In terms of physical $CNOT$ cost, the comparison also is favorable. To analyze a specific example of the $3k+8$-to-$k$ scheme which has been optimized, the 20-to-4 magic state distillation circuit has 61 logical $CNOT$ gates\cite{Litinski2019magicstate}. Each of these logical $CNOT$ gates is 23 physical $CNOT$ gates, for a total of 1403 physical $CNOT$ gates. Considering a large number of distillation procedures running in parallel, for every 1 high-fidelity encoded magic state that is output, this procedure needs 25 noisy magic states and two rounds of distillation. The first round has 5 times as many magic state distillations involved, and so needs 5 times as many $CNOT$ gates. This results in a net physical $CNOT$ cost of $1403 \times 5 + 1403 = 8418$. This is without including any error correction costs. In comparison, assuming pessimistically, again, that 12 noisy ancilla are needed for the doubled code to produce a single high-fidelity $\ket{+}$ state, and assuming even further pessimistically a physical $CNOT$ cost of 473 per noisy ancilla, obtained from Steane's basic Latin rectangle method\cite{steane2002codewords} for preparing a $\ket{+}$ state\footnote{The cost for the $\ket{0}$ state is only 359 even for the basic Latin rectangle method, and exploiting the overlap, circuits that reduce the $CNOT$ cost for preparing noisy $\ket{0}$ and $\ket{+}$ ancilla to 267 and 315, respectively, were found.}, the noisy ancilla preparation phase would require 5676 physical $CNOT$ gates. The 12 noisy ancilla would then involve 11 logical $CNOT$ gates, for $95 \times 11 = 1045$ additional physical $CNOT$ gates. This puts the final $CNOT$ cost comparison per high fidelity magic state at 8418 for the 20-to-4 magic state distillation procedure and 6721 for the doubled code procedure. To reiterate, this is comparing an optimistic gate count for magic state distillation and a pessimistic, unoptimized gate count for the doubled code procedure, and the comparison is already favorable.

Optimization of the ancilla preparation could potentially have a big impact for the doubled code procedure. It could potentially drastically lower the physical $CNOT$ and qubit cost for Steane-style ancilla preparation. Optimization would also be important for maximizing the error threshold of the procedure. However, even when unoptimized, the resource costs, both with $CNOT$ count and qubit count, are competitive with magic state distillation applied to the quantum Golay code. A benefit of magic state distillation procedures is the higher error threshold; which scheme performs better could depend on the physical error rate of a given device.

\section{Conclusions}
\label{sec:conc}
Achieving fault-tolerant quantum computation is a requirement for scalable quantum computers. Although different avenues are possible for fault tolerance, code concatenation schemes remain a promising approach. The quantum Golay code has been recognized as a promising candidate for this purpose, but the existing methods to achieve a universal gate set for this code in the literature are based on magic states and magic state distillation. In this paper, a triorthogonal code was constructed that can be related to the quantum Golay code. This construction enables a simple code conversion procedure for the quantum Golay code which enables the fault-tolerant implementation of a universal gate set without the use of magic state distillation, and resource costs measured in $CNOT$ gate count and physical qubit count are competitive with magic state distillation schemes for the Golay code. Code conversion can be performed using transversal $CNOT$ gates with an encoded ancilla rather than the previous methods of measuring stabilizers. The new code, with its transversal $T$ gate, could also used for directly preparing magic states encoded in the Golay code, allowing the usual magic state techniques without the need for distillation. The constructed triorthogonal code could also be used as a triorthogonal quantum error-correcting code for other purposes. The procedure generalizes to a wide class of self-dual CSS codes and can be used to construct related families of $k$-orthogonal codes analogous to the Reed-Muller code family.

\section*{Acknowledgements}
MS is supported by the United States Department of Energy under Grant Contract DE-SC0012704.

\bibliography{references.bib}

\end{document}